\documentclass[a4paper]{jpconf}
\usepackage{graphicx}
\usepackage{psfig}

\def\sgra{Sgr~A$^*$}

\begin{document}
\title{Sgr~A$^*$: The Optimal Testbed of Strong-Field Gravity}

\author{Dimitrios Psaltis and Tim Johannsen}

\address{Astronomy and Physics Departments, The University of Arizona,
933 N.\ Cherry Ave, Tucson, AZ 85721, USA}

\ead{dpsaltis@email.arizona.edu, timj@physics.arizona.edu}

\begin{abstract}
The black hole in the center of the Milky Way has been observed and
modeled intensely during the last decades. It is also the prime target
of a number of new experiments that aim to zoom into the vicinity of
its horizon and reveal the inner working of its spacetime. In this
review we discuss our current understanding of the gravitational field
of \sgra\ and the prospects of testing the Kerr nature of its
spacetime via imaging, astrometric, and timing observations.
\end{abstract}

\section{Introduction}

The Kerr spacetime of spinning black holes is one of the most
unexpected predictions of Einstein's theory of General Relativity. The
special role this spacetime plays in the theory of gravity is
encapsulated in the no-hair theorem, which states that the Kerr metric
is the only stationary, axisymmetric, asymptotically flat solution to
the vacuum field equations that possesses a horizon but no closed
timelike loops (Israel 1967, 1968; Carter 1971, 1973; Hawking 1972;
Robinson 1975; Mazur 1982). Because of this theorem, it is expected
that all astrophysical objects that have been identified as black-hole
candidates are indeed described by the Kerr metric.

The Kerr nature of astrophysical black holes is a prediction that can
be tested observationally. Doing so offers the unique opportunity of
both rejecting alternative interpretations of their nature and of
verifying General Relativity in the strong-field regime.  There are at
least three astrophysical settings we know of today, in which black
holes appear to exist: in galactic X-ray binary systems, as
ultraluminous X-ray sources, and in active galactic nuclei. Among all
these black holes, the one in the center of our galaxy, \sgra,
combines a large mass with a small distance from the sun (e.g., Ghez
et al.\ 2008; Gillessen et al.\ 2009). It can also be probed by a
variety of observations throughout the electromagnetic spectrum making
it the optimal test bed of the Kerr metric.

Recent and anticipated advances in observations of \sgra\ throughout
the electromagnetic spectrum have both secured our understanding of
the basic properties of this black hole (e.g., Reid 2009), and opened
new opportunities for devising tests of gravity theories.  At the same
time, there have been significant recent advances in the development
of a theoretical framework with which observations of black holes can
be used to test quantitatively the Kerr metric and search for
violations of the no-hair theorem.
 
In this article, we review our current understanding of the properties
of the black hole in the center of the Milky Way and discuss the prospect
of testing the no-hair theorem with upcoming observations. 

\begin{figure}[t]
\centerline{\psfig{file=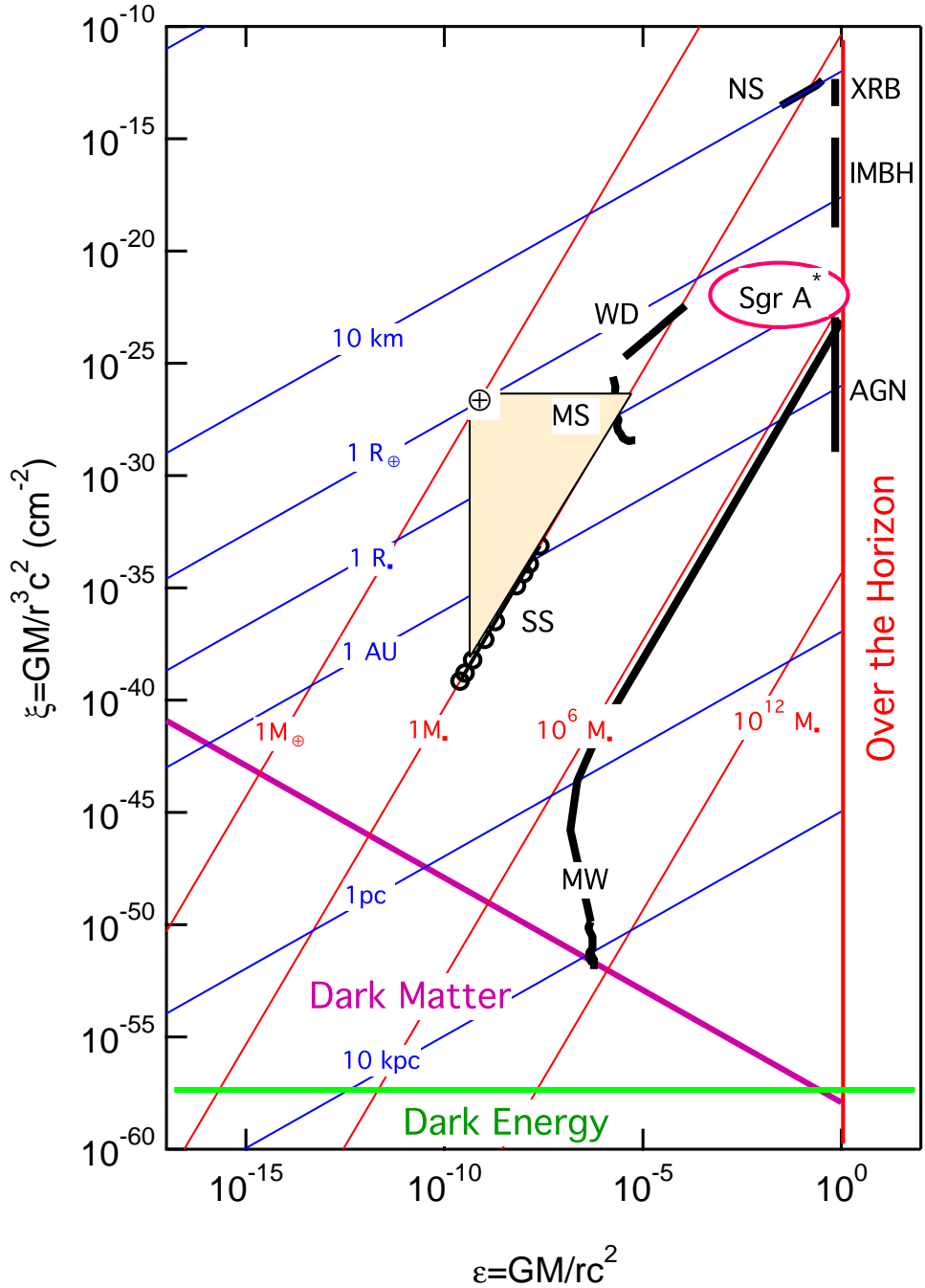,width=5.5in}}
\caption{\footnotesize A parameter space for tests of General
  Relativity.  The $x-$axis measures the potential and the $y-$axis
  the curvature of the field probed by different experiments. A number
  of astrophysical and cosmological objects are also shown. The black
  hole in the center of the Milky Way, \sgra, probes a region of the
  parameter space that has not been investigated with other current
  tests; the location of the latter is outlined by the inverted yellow
  triangle in the center of the figure (Psaltis 2008).}
\label{fig:tests}
\end{figure}

\section{Sgr A*: The Optimal Black Hole for Tests of the No-Hair Theorem}

\begin{figure}[t]
\centerline{\psfig{file=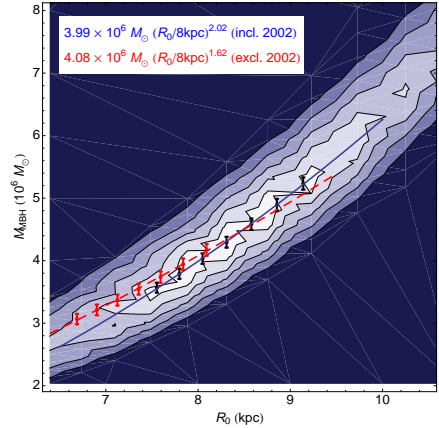,width=2.3in}}
\caption{\footnotesize Confidence contours of the mass and distance
to \sgra, based on the analysis of orbits of nearby stars (Gillessen et al.\
2009).}
\label{fig:mass}
\end{figure}

\begin{figure}[t]
\centerline{\psfig{file=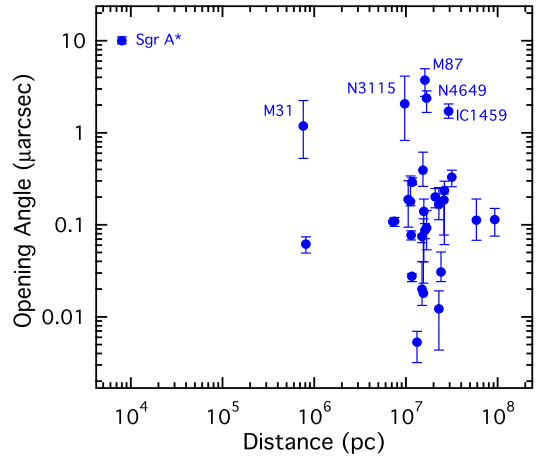,width=3in}}
\caption{\footnotesize The opening angles, as viewed by
  an observer on Earth, of the horizons of a number of supermassive
  black holes in distant galaxies with a secure dynamical mass
  measurement (sample of Tremaine et al.\ 2002). The opening angle of
  the black hole horizon in the center of the Milky Way (Sgr~A$^*$) is
  also shown for comparison (after Psaltis 2008).}
\label{fig:images}
\end{figure}

The black hole in the center of the Milky Way is the optimal
astrophysical object for testing our understanding of black hole
spacetimes, for a number of reasons that we outline below.

\begin{figure}[t]
\centerline{\psfig{file=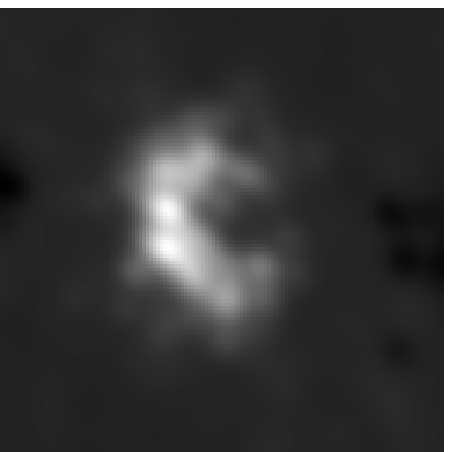,width=2.in}
\psfig{file=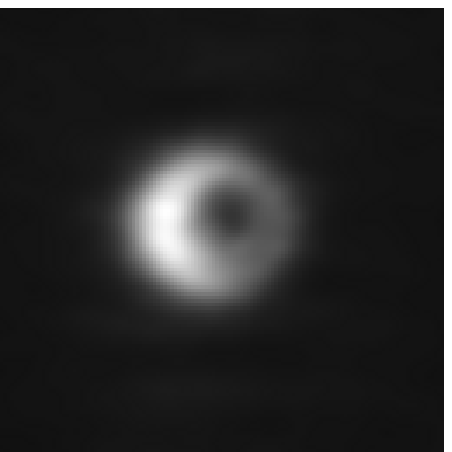,width=2.in}}
\caption{\footnotesize Simulated images of the inner accretion flow
  around \sgra, as they will be resolved with {\em (left)\/} a
  7-telescope array in the near future (3-5 years) and {\em (right)\/}
  a 13-telescope array (Fish \& Doeleman 2009).}
\label{fig:inter}
\end{figure}

\begin{itemize}

\item{{\bf \sgra\ probes a new regime of gravitational field
    strengths.}  General Relativity has been tested in a rather
  limited range of gravitational field strengths
  (Fig.~\ref{fig:tests}). Phenomena in the vicinity of the black-hole
  horizon around \sgra\ probe gravitational fields with a potential
  and a curvature that are five orders of magnitude larger than
  the strongest fields probed by current tests of General Relativity.}

\item{{\bf \sgra\ has an accurately measured mass}. Fitting the
  trajectories of stars in the vicinity of \sgra\ has lead to an
  accurate measurement of its mass. A recent careful analysis of the
  orbits by Gillessen et al.\ (2009; see also Ghez et al.\ 2008)
  resulted in a measurement of $M=(4.31\pm 0.06 \pm 0.36)\times 10^6
  M_\odot$ with the second uncertainty corresponding to the error in
  the estimate of the distance to the center of the Milky Way (see
  Fig.~\ref{fig:mass}). This highly accurate mass measurement
  specifies one of the only two basic parameters of the black hole and
  sets uniquely the scale for any anticipated relativistic phenomena.}

\item{{\bf The mass measurement provides an independent distance
    measurement to \sgra}.  Testing the Kerr spacetime via, e.g.,
  analysis of the images of its inner accretion flow, requires an
  independent knowledge of the distance to the source in order to
  convert angular separations to physical lengths. Studies of the orbits
  of stars that approach \sgra\ lead to a measurement of the distance
  to the source of $R_0=8.33\pm 0.35$~kpc (Gillessen et al.\ 2009).}

\item{{\bf \sgra\ is one of the only two black-holes with horizons
    that will be resolved with sub-mm VLBI in the very near future.}
  For a supermassive black hole in a distant galaxy, the opening angle
  of the horizon as viewed from the earth is
\begin{equation}
\theta=20\left(\frac{M}{10^9\,M_\odot}\right)
\left(\frac{1~\mbox{Mpc}}{D}\right)~\mu\mbox{arcsec}\;.
\end{equation}
This is shown in Figure~\ref{fig:images} for a number of supermassive
black holes with secure mass determinations. \sgra\ is the black hole
that combines the highest brightness with the largest angular size of
the horizon.}

\begin{figure}[t]
\centerline{\psfig{file=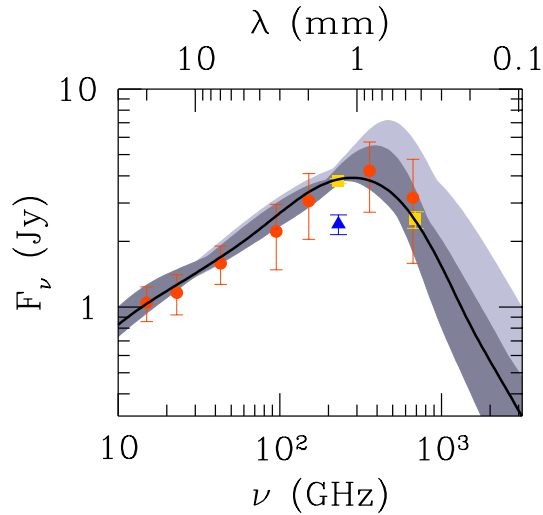,width=2.8in}}
\caption{\footnotesize The long-wavelength spectrum of \sgra. The grey
  filled areas show the envelopes of semi-analytic models that best
  match the data (Broderick et al.\ 2009).}
\label{fig:spectrum}
\end{figure}

\begin{figure}[t]
\centerline{\psfig{file=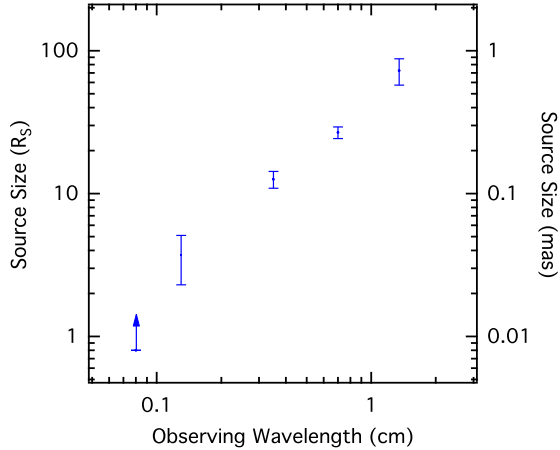,width=3in}}
\caption{\footnotesize The major axis of the accretion flow around the
  black hole in the center of the Milky Way, as measured at different
  wavelengths, in units of the Schwarzschild radius (left axis) and in
  milliarcsec (right axis; adapted from Shen et al. 2005 and Doeleman
  et al. 2008; after Psaltis 2008). Even with current technology, the
  innermost radii of the accretion flow can be readily observed.}
\label{fig:size}
\end{figure}

\item{{\bf At photon frequencies $> 5\times 10^{11}$~Hz, the emission from
    \sgra\ is optically thin.}  The long-wavelength spectrum of
  Sgr~A$^*$ peaks at a frequency of $\simeq 5\times 10^{11}$~Hz, suggesting
  that the emission changes from optically thick (probably synchrotron
  emission) to optically thin at a comparable frequency (see
  Fig.~\ref{fig:spectrum}). As a result, observations at frequencies
  comparable to or higher than this transition frequency are expected
  to probe the region close to the black-hole horizon without
  significant obscuration. This fact is indeed confirmed by the
  measurement of the size of \sgra\ at this part of the spectrum (see
  below).}

\item{{\bf The size of the emitting region in \sgra\ at long
    wavelengths is comparable to the horizon scale.} Since the first
  measurements of the size of the source at 7~mm and at 1.4~mm
  (Krichbaum et al.\ 1998) demonstrated that the emitting region is
  only a few times larger than the radius of the horizon (see
  Figure~\ref{fig:size}), a number of observational investigations
  have aimed to probe deeper into the gravitational field of the black
  hole. Most recently, observations at 1.3~mm (Doeleman et al.\ 2008)
  revealed near-horizon sizes for the emitting region of \sgra.}

\item{{\bf The detailed structure of the inner accretion flow around
    \sgra\ will be imaged in the very near future.}
  Figure~\ref{fig:inter} shows two simulated images of \sgra\ at
  345~GHz as they will be resolved by a 7-telescope array in the near
  future (3-5 years) and by a 13-telescope array further in the future
  (Fish \& Doeleman 2009). The characteristic asymmetry in the image
  caused by Lorentz boosting as well as the shadow of the black hole
  are clearly visible.}

\end{itemize}

\section{Parametrizing Deviations from the Kerr Metric}

The Kerr metric, which describes the exterior spacetime of a spinning
black hole, is not unique to General Relativity. Indeed, a large
number of simple extensions to the general relativistic field
equations are also satisfied by the Kerr solution (Psaltis et
al.\ 2008). This is perhaps not surprising (see, e.g., Barausse \&
Sotiriou 2008), given the fact that the Kerr metric is derived by
solving the vacuum field equation
\begin{equation}
R_{\mu\nu}=0\;,
\label{eq:vacuum}
\end{equation}
where $R_{\mu\nu}$ is the Ricci tensor, and not the full Einstein
field equation (the two are equivalent in
vacuum). Equation~(\ref{eq:vacuum}), however, is also satisfied by the
Minkowski metric of special relativity. As a result, the field
equations of any theory that has special relativity as its vacuum
limit should also be equivalent in vacuum to equation~(\ref{eq:vacuum}) 
and, therefore, be satisfied by the Kerr metric. This argument remains
valid even when we allow for the possibility of a finite cosmological
constant (see Psaltis et al.\ 2008).

Constructing a modification to General Relativity in which the black
hole solutions are not described by the Kerr metric can be done, in
principle, by violating one of the more fundamental assumptions in our
understanding of gravity, such as the equivalence principle (see,
e.g., Eling \& Jacobson 2006). The only black-hole solution in a theory that
obeys the equivalence principle that we are aware of is that of Yunes
\& Pretorius (2009) and Konno et al. (2009) in Chern-Simons gravity.
In Chern-Simons gravity, a parity violating terms is added to the
Einstein-Hilbert action of General Relativity, with a coupling
constant that is a dynamical field (see Alexander \& Yunes 2009 for a
review). This parity violating term is zero for a parity-symmetric
metric and, therefore, the Minkowski and Schwarzschild spacetimes
remain solutions to the field equations. However, the spacetime of a
spinning black hole has a particular parity specified by its spin and,
therefore, it is different than the Kerr metric.

Given the very limited range of available black-hole metrics in
modified gravity theories that is known today, it is more fruitful to
test the Kerr metric in a phenomenologial way, by adding
parametrically terms to its various elements (see Psaltis 2009 for a
discussion). This is equivalent to testing for violations of the no-hair
theorem, even within General Relativity.

The no-hair theorem states that the only stationary, axisymmetric,
asymptotically flat vacuum solution to the Einstein field equations
that possesses a horizon and no time-like loops is the Kerr metric (we
do not consider here the unlikely possibility that an astrophysical
black hole will have a net charge). The Kerr metric is uniquely
determined by only two parameters, the mass and the spin of the black
hole. This allows us to define a formal test of the no-hair theorem,
based on the work of Ryan (1995), in the following way (see also
Collins \& Hughes 2004; Glampedakis \& Babak 2006; Vigeland \& Hughes
2010; Vigeland 2010).

We can, in principle, expand the exterior metric of any compact object
in multipoles and use observations to measure the coefficients of the
expansion. Because of the no-hair theorem, only two of the multipole
coefficients for the spacetime of a black hole are independent. The
coefficient of the monopole is the mass $M$ of the black hole and of
the dipole is its spin $a$. All higher-order coefficients will depend
on the first two, in the particular way dictated by the Kerr
metric. Testing the no-hair theorem requires measuring at least the
coefficient of the quadrupole $q$ and verifying whether it satisfies
the Kerr relation $q=-a^2$.

Four different approaches have been explored so far for introducing
additional non-Kerr hair to the spacetimes of compact objects. Ryan
(1995) studied a general expansion of stationary, axisymmetric,
asymptotically flat spacetimes in Geroch-Hansen multipoles. Collins \&
Hughes (2004) as well as Vigeland \& Hughes (2010) added Weyl-sector
bumps to the Schwarzschild and Kerr spacetimes. Glampedakis \& Babak
(2006) modified the Kerr metric by adding an arbitrary quadrupole
moment while ensuring that the metric to that order remains a solution
to the vacuum field equation~(\ref{eq:vacuum}). Finally, Gair et
al. (2008) explored in detail the parametric solution of Manko \&
Novikov (1992), which allows for deviations of all higher-order
multipoles from their Kerr values. All these approaches were developed
originally in order to test General Relativity with future
observations of the gravitational waves generated during inspirals
into supermassive black holes. However, they are also directly
applicable to tests of gravity with \sgra.

In Johannsen \& Psaltis (2010a, 2010b, 2010c), we focused 
on the parametric post-Kerr spacetime obtained by Glampedakis \&
Babak (2006) for two reasons. First, this approach uses a single
parameter associated to the quadrupole moment of the spacetime to
quantify potential deviations from the Kerr metric, making it the
simplest and most concise possible avenue to testing the no-hair
theorem. Second, the complete metric of Glampedakis \& Babak (2006)
remains a valid solution to the vacuum Einstein field equations,
allowing us to perform a self-consistent test of the no hair theorem and of
the black-hole identification of the compact object, within General
Relativity. 

\begin{figure}[t]
\centerline{\psfig{file=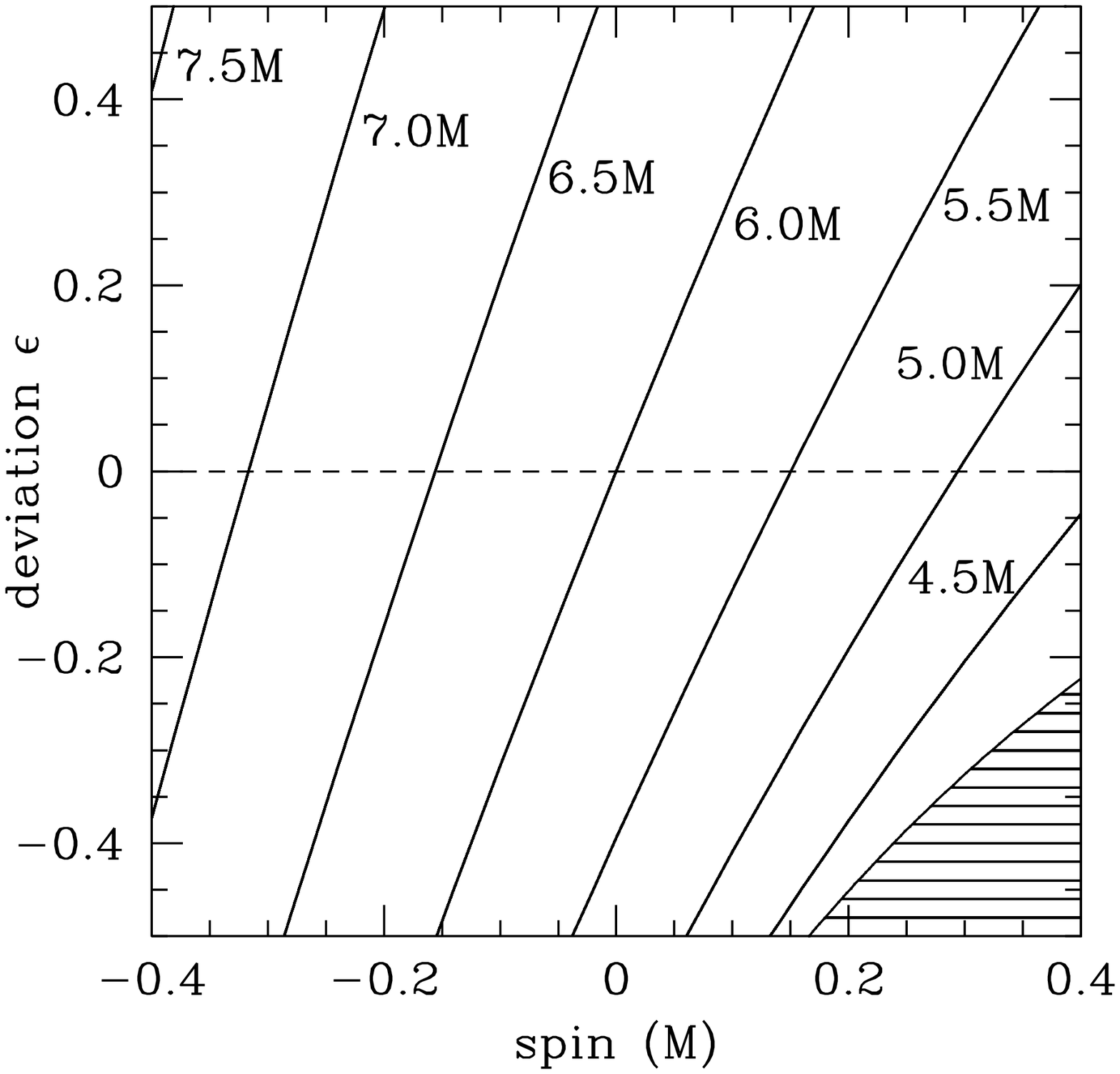,width=2.8in}
\psfig{file=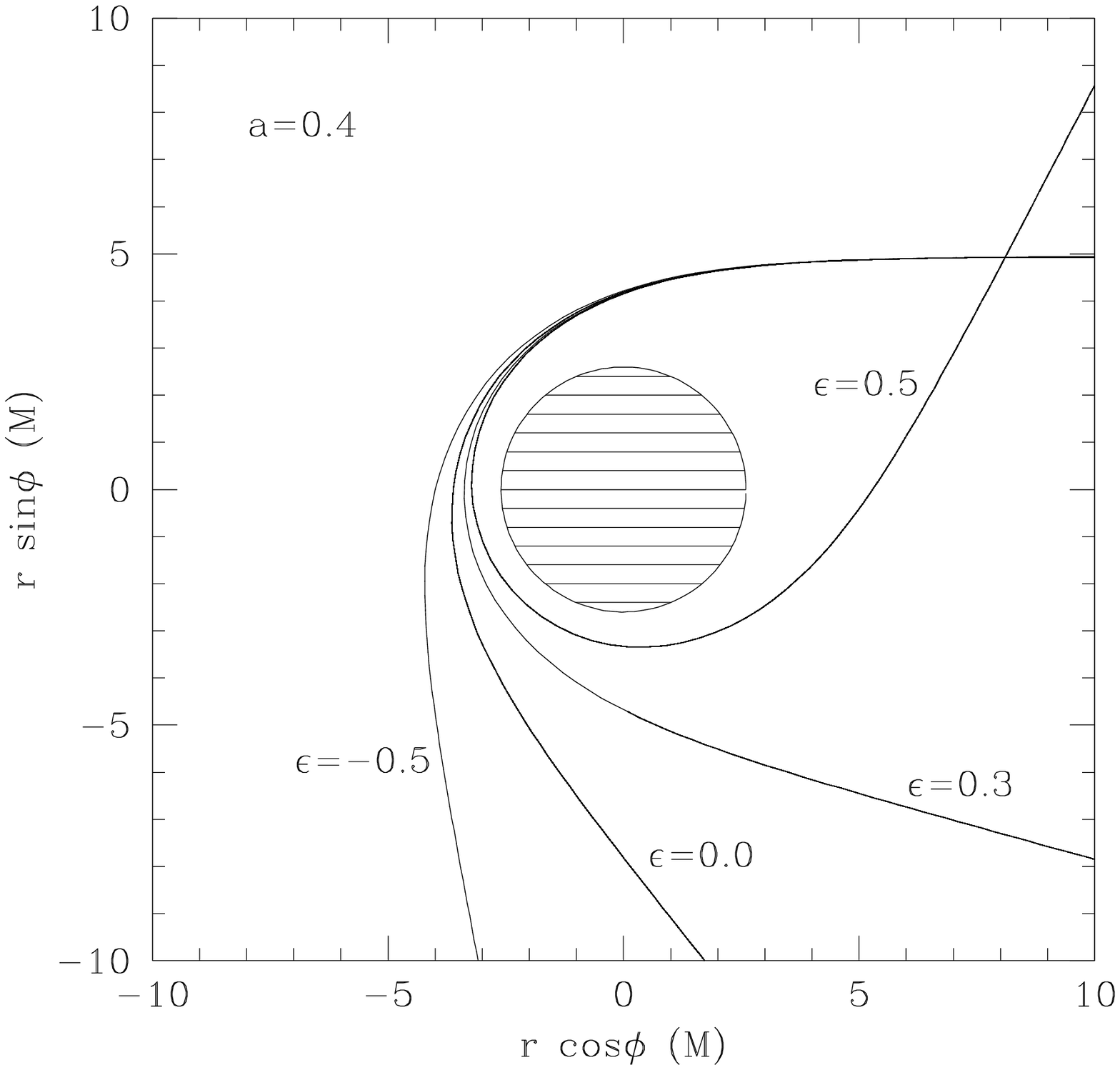,width=2.8in}}
\caption{\footnotesize The dependence of {\em (Left)\/} the radius
  of the innermost stable circular orbit around a black hole and of
  {\em (Right)} the trajectories of photons on the black-hole spin $a$
  and on the parameter $\epsilon$ that measures the degree of
  violation of the no-hair theorem (Johannsen \& Psaltis 2010a, 2010c).}
\label{fig:quadrupole}
\end{figure}

The key drawback of all currently proposed parametric deviations from
the Kerr metric is that they cannot be used in describing the exterior
spacetimes of rapidly spinning black holes. Precisely because of the
no-hair theorem, it is impossible to construct a spacetime that
deviates from the Kerr metric, is a solution to the general
relativistic vacuum field equations, and is regular everywhere outside
the black-hole horizon; such a construction would prove that the
no-hair theorem is violated. In all parametric deviations from the
Kerr metric, the spacetimes become irregular at radii $\simeq 2M$ (see
Gair et al.\ 2008; Johannsen \& Psaltis 2010a). For moderate
black-hole spins, radii comparable to $2M$ are much smaller than the
radius of the photon orbit and of the innermost stable circular
orbit. As a result, such radii can be artificially excised, without
affecting the prediction of any observable. For rapidly spinning black
holes, however, all characteristic radii become $\le 2M$ and,
therefore, the irregularities of the spacetimes preclude us from
calculating various observables related, e.g., to the inner accretion
flow around a black hole. We do not anticipate this to introduce
significant problems for the case of \sgra, because preliminary
attempts to model the VLBI images from the source show preference for
small values of its spin ($a\le 0.3$; Broderick et al.\ 2009, 2010).

\section{Testing the No-Hair Theorem with Sgr~A$^*$}

In Johannsen \& Psaltis (2010a), we explored in detail the properties
of the Glampedakis \& Babak (2006) metric, addressing its potential in
testing the no-hair theorem with astrophysical observations in the
electromagnetic spectrum. Following their original analysis, we
expressed the coefficient of the quadrupole multipole of the spacetime
as
\begin{equation}
q=-(a^2+\epsilon)
\end{equation}
with the parameter $\epsilon$ measuring the degree of violation of the
no-hair theorem. We then studied the trajectories of photons and
particles in this spacetime and identified three important effects of
the presence of a non-Kerr quadrupole. 

First, the location of the innermost stable circular orbit (ISCO) is
significantly altered, as shown in Figure~\ref{fig:quadrupole} (left
panel). The location of the maximum emission from the accretion flow
around \sgra\ is expected to be very close to that of the ISCO (see,
e.g., Krolik \& Hawley 2002 and references therein) and, therefore,
the brightness profile of the image from \sgra\ will depend on the
value of the quadrupole. Second, the radius of the photon orbit is
also significantly affected by the value of the quadrupole. The radius
of the photon orbit determines the size of the shadow of the black
hole (see Bardeen 1973; Falcke et al.\ 2000), and generates one of the
most prominent features in the image of \sgra\ (i.e., the structure
that we call the photon ring; see below). Finally, the detailed
trajectories of photons that propagate close to the photon orbit
depend also on the magnitude of the quadrupole (see
Fig.~\ref{fig:quadrupole}, right panel) leading to non-trivial
deformations of the calculated images.

Figure~\ref{fig:quadrupole} also shows, that changing the quadrupole
moment of a black-hole spacetime may even alter qualitatively the
properties of particle and photon orbits in the vicinity of its
horizon.  In the Glampedakis \& Babak (2006) metric, for each value of
the spin, there exists a minimum quadrupole deviation beyond which all
circular orbits are stable to radial perturbations (Johannsen \&
Psaltis 2010c; this results has also been discussed for the Manko \&
Novikov metric by Gair et al.\ 2008). For relatively large black-hole
spins ($a>0.4$), any reduction of the quadrupole moment of the
spacetime from its Kerr value stabilizes all circular orbits against
radial perturbations. In those spacetimes, equatorial particle orbits
close to the black-hole horizon become instead unstable to vertical
perturbations (see Gair et al.\ 2008; Johannsen \& Psaltis 2010c) with
implications for the inner accretion flows that have not been explored
yet.

\begin{figure}[t]
\centerline{\psfig{file=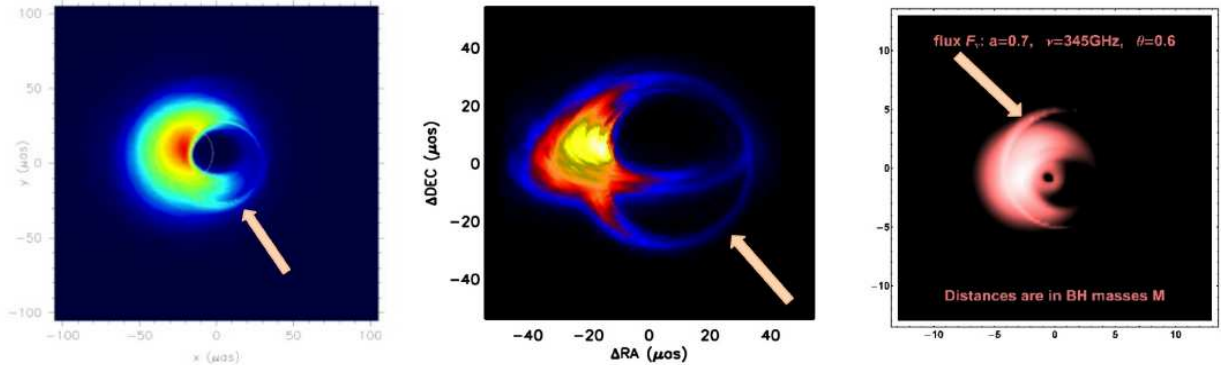,width=6.5in}}
\caption{\footnotesize The presence of a bright photon ring
  surrounding the black-hole shadow is ubiquitous in all current
  simulations of images from radiatively inefficient accretion flows
  (from left: Moscibrodzka et al. 2009; Dexter et al.\ 2009;
  Shcherbakov \& Penna 2010). The size, location, and shape of the
  photon ring is determined by the projected photon orbit at infinity
  and, as such, it depends only on the metric of the spacetime. The
  accretion flow is necessary only to provide the photons that will
  trace the photon ring for an observer at infinity.}
\label{fig:ring}
\end{figure}

In Johannsen \& Psaltis (2010b) we investigated the prospect of
measuring the parameter $\epsilon$ for Sgr~A$^*$ via imaging
observations in the sub-mm. We identified in the simulated images of
the accretion flow a bright ring (which we called the photon ring)
that surrounds the black-hole horizon (see also, e.g., Beckwith \& Done
2005). This ring is produced by photon rays that orbit the black hole
multiple times at the radius of the photon orbit before escaping to
infinity. Its presence is ubiquitous in all current simulations of
images from radiatively inefficient accretion flows (see
Fig.~\ref{fig:ring}).

\begin{figure}[t]
\centerline{\psfig{file=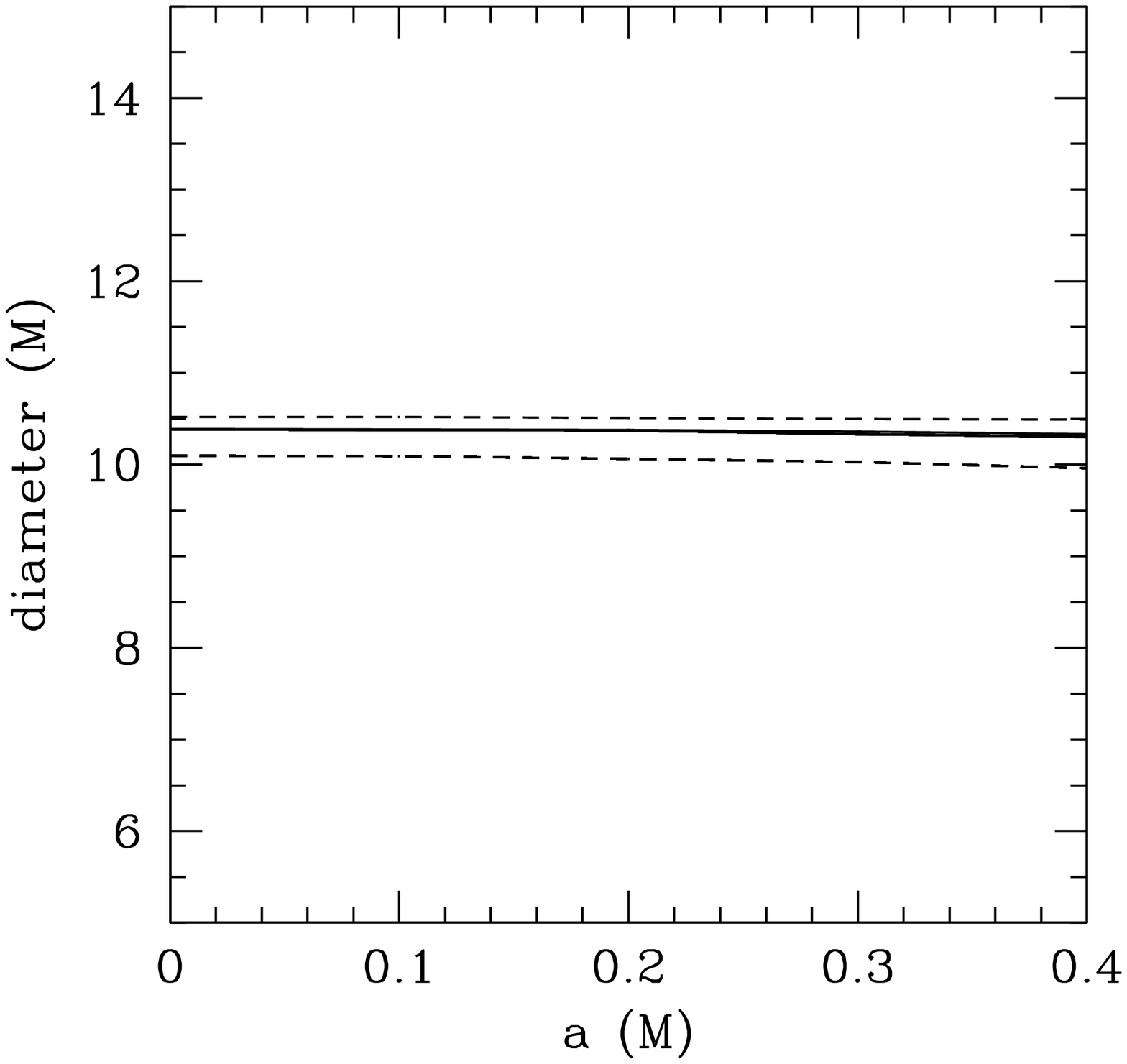,width=2.2in}
\psfig{file=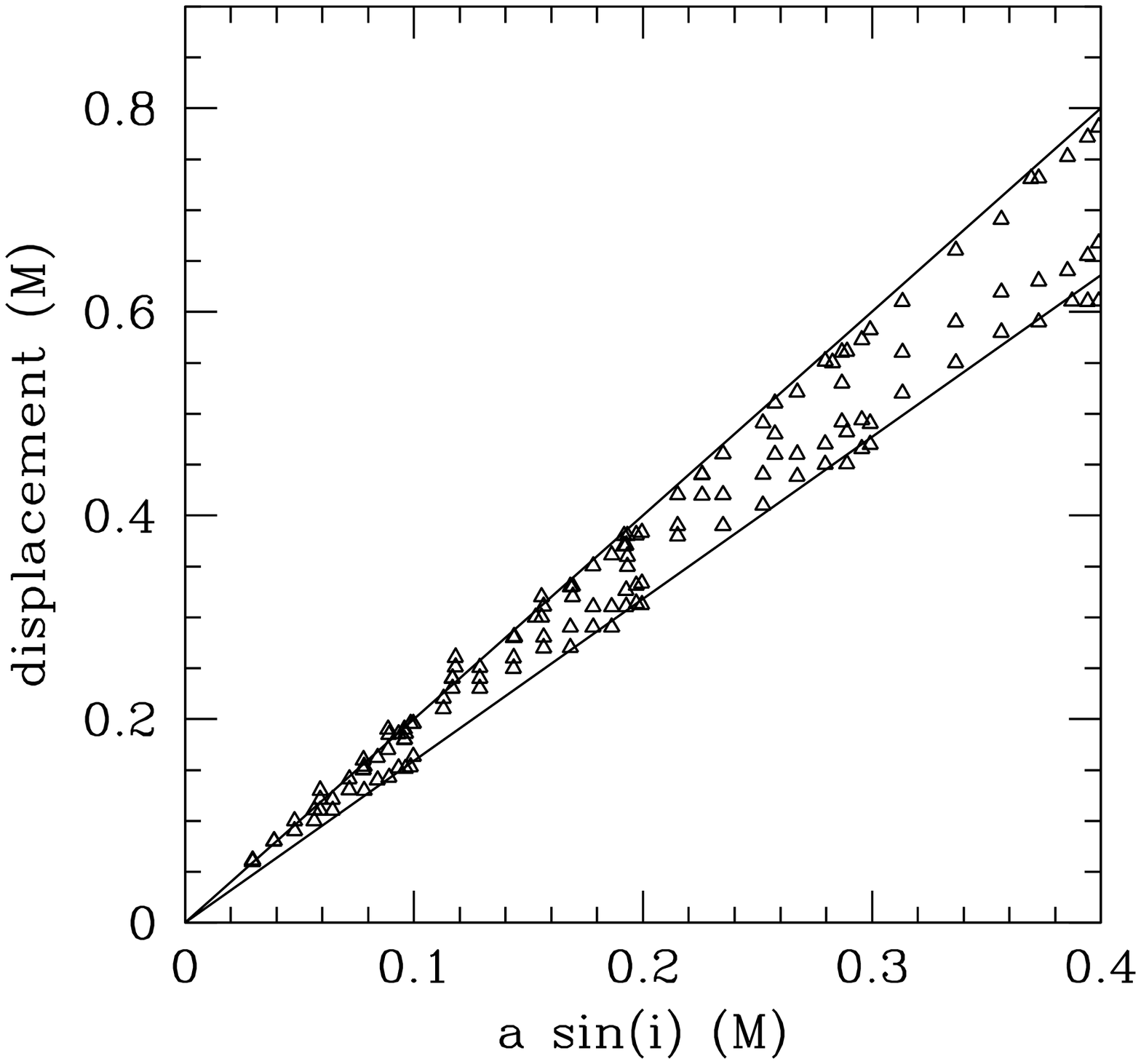,width=2.2in}
\psfig{file=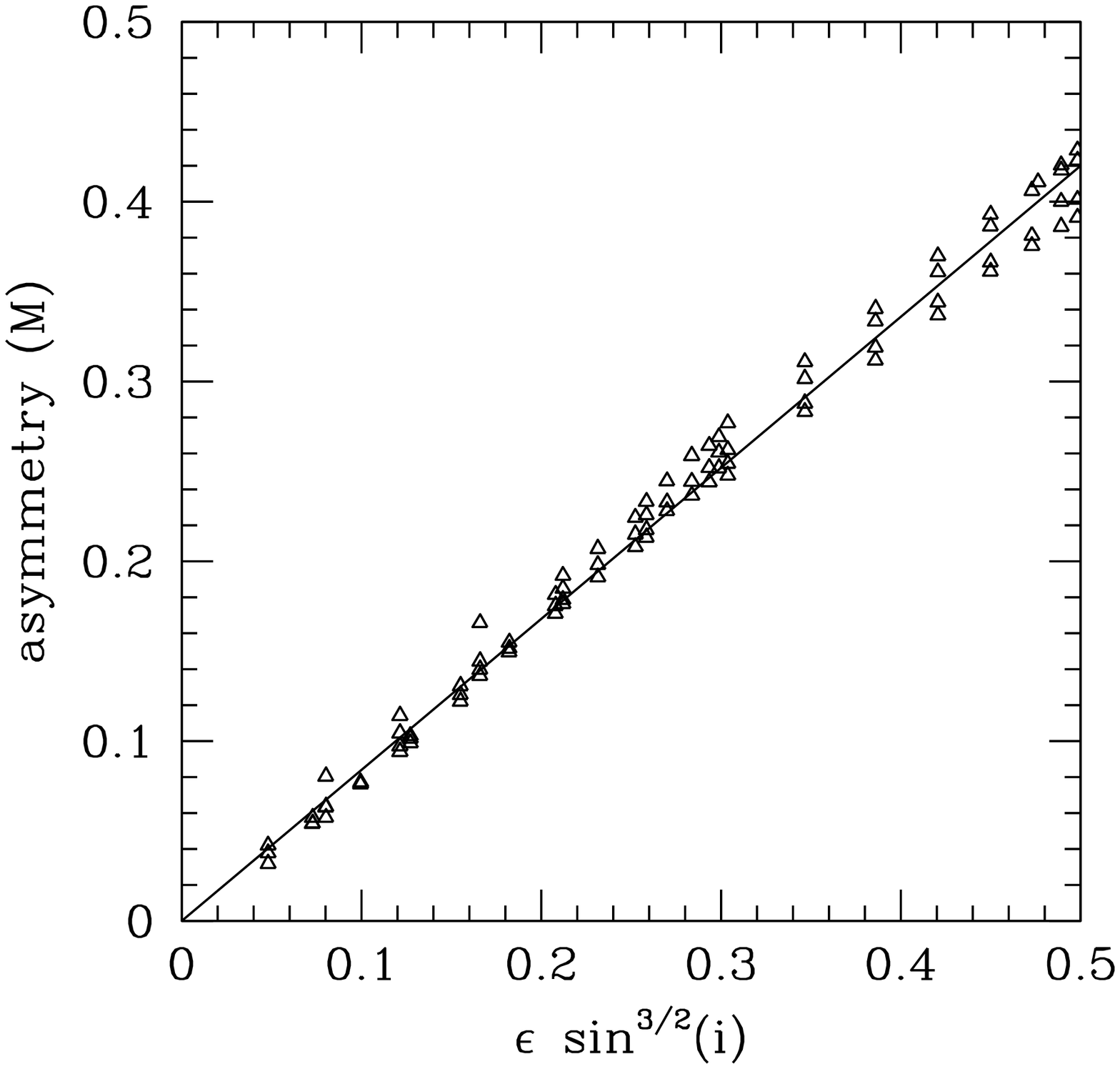,width=2.2in}}
\caption{\footnotesize Three panels showing the {\em (Left)\/}
  diameter, {\em (Middle)\/} displacement, and {\em (Right)\/}
  asymmetry of the bright photon ring surrounding the black-hole
  shadow in the image of an accreting black hole, as a function of the
  parameters of the quasi-Kerr metric. The three depicted properties
  of the ring measure three different multipoles of the spacetime
  (modulo the sine of the observer's inclination $\sin i$) and can
  lead to a test of the no-hair theorem (Johannsen \& Psaltis 2010b).}
\label{fig:param_images}
\end{figure}

The location, size, and geometry of the photon ring corresponds to the
projection of the photon orbit on the image plane of the distant
observer. Because of this, its geometric characteristics are
independent of the particular accretion flow model assumed.  The
accretion flow is necessary only to provide the photons that will
trace the photon ring for an observer at infinity. The diameter of the
bright photon ring is $\sim 10.0-10.5~M$, practically independent of
the spin and the quadrupole moment of the black hole, providing a
direct measure of the black-hole mass (Figure~\ref{fig:param_images}).
The displacement of the ring from the center of mass of the system
depends primarily on the spin of the black hole. Finally, the
deviation of the shape of the photon ring from a circle is a direct
measure of the violation of the no-hair theorem (modulo the observer's
inclination). Indeed, photon rings for the Kerr metric remain
practically circular for all but the fastest spins ($a\le 0.9$),
whereas even a small degree of violation of the no-hair theorem (as
measured by the parameter $\epsilon$) introduces a significant
asymmetry to the photon ring (see
Figure~\ref{fig:param_images}). Imaging the inner accretion flow and
measuring the asymmetry of the photon ring offers, therefore, the
possibility of a clean, quantitative test of the no-hair theorem.

\section{Distinguishing Gravitational Effects from Astrophysical
Complications}

The biggest challenge in searching for violations of known physics
with astrophysical measurements is ensuring that a particular
measurement is not affected adversely by astrophysical complications.
In the previous section, we identified a signature of gravitational
effects that probes cleanly the metric of the black hole and is
affected only marginally by other astrophysics, i.e., the bright
photon ring that surrounds the shadow of the black hole.  The
location, size, and shape of the photon ring depends predominantly on
the spin and quadrupole of the spacetime and only marginally on the
properties of the underlying accretion flow. Even in this case,
however, the narrow width of the ring (of order a few tens of $M$), as
well as possible confusion with other emission from the accretion flow
between the observer and the horizon may make such a measurement
biased.

We can convert a compelling argument for possible violations of the
no-hair theorem into a bullet proof result by using different
observational probes for the spin and quadrupole moment of the same
black hole and by testing whether all such probes agree
quantitatively. \sgra\ is unique in that it potentially offers two
additional, independent ways of testing the no-hair theorem.

\begin{figure}[t]
\centerline{\psfig{file=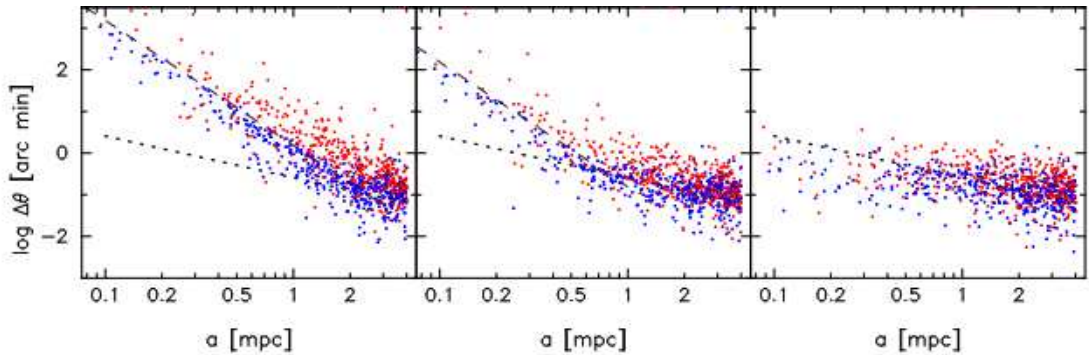,width=6.0in}}
\caption{\footnotesize The cummulative change in the angle $\delta
  \theta$ between the initial and final orbital angular momenta of
  stars around \sgra\ with different orbital semi-major axes $a$. The
  dashed lines show the effect of relativistic frame-dragging while
  the dotted lines correspond to an approximate model of precession due
  to stellar perturbations. The three panels show the result for black
  hole spins {\em (left)\/} $a=1$, {\em (middle)\/} $a=0.1$, and {\em
    (right)\/} $a=0$ (Merritt et al.\ 2010).}
\label{fig:prec}
\end{figure}

Will (2008) showed that astrometric observations of the orbits of
stars clustered very close to the black hole can be used to measure
the quadrupole moment of the black hole via detection of frame
dragging. In a follow-up study, Merritt et al. (2010) showed that
interactions between the stars of the cluster will mask frame dragging
effects, unless the analysis is confined to stars within $\simeq 1000$
Schwarzschild radii from the horizon (see Fig.~\ref{fig:prec}). If
such stars exist, future instruments (such as GRAVITY; Bartko et
al.\ 2009) will be able to provide an independent measurement of the
quadrupole moment of the black-hole spacetime.

\begin{figure}[t]
\centerline{\psfig{file=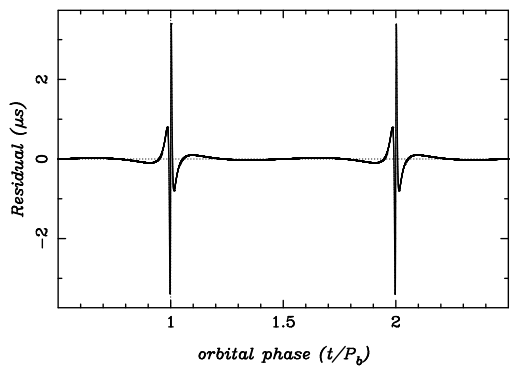,angle=-90,width=3.0in}}
\caption{\footnotesize Typical residuals in the timing solution for a
  radio pulsar orbiting a black hole, when the effects of the
  quadrupole moment of the black hole spacetime have not been taken
  into account (Wex \& Kopeikin 1999). The residuals shown here have
  been calculated for a $10^4 M_\odot$ black hole; they will be
  similar in shape but larger in magnitude for a pulsar orbiting
  \sgra.}
\label{fig:psr}
\end{figure}

The presence of a cluster of massive stars around Sgr~A$^*$ makes it
practically certain that a number of radio pulsars will also be
orbiting the black hole. Current surveys put an upper limit of as many
as 90 normal pulsars within the central parsec of the Galaxy (e.g.,
Macquart et al.\ 2010). Detecting such a pulsar in one of the current
surveys will allow us to measure the quadrupole moment of the black
hole by looking for particular spin-orbit coupling residuals in the
timing solution for each pulsar (see Fig.~\ref{fig:psr}; Wex \&
Kopeikin 1999).

The combination of observations of the images from the inner accretion
flow around \sgra, of the relativistic precession of massive stars
orbiting close to the black hole, and of the residuals in the timing
solution of radio pulsars in similar orbits will allow for three,
independent measurements of the spin and of the quadrupole moment of
the black hole. Each observation will be performed at different
wavelengths and with different telescopes. Moreover, each of the three
observed phenomena will probe different regions of the black-hole
spacetime and will be affected by different systematics. If all three
measurements give consistent solutions for the black-hole spin and
quadrupole moment, then this will lead to an unequivocal test of the
no-hair theorem with an astrophysical black hole.

\end{document}